\documentclass[11pt]{article}

\usepackage{geometry}  
\usepackage[T1]{fontenc}  
\usepackage[utf8x]{inputenc}  
\usepackage{lineno}  
\usepackage{fancyhdr}  
\usepackage{enumitem}  
\usepackage{hyperref}  
\usepackage{titling}  
\usepackage[square,numbers,sort&compress]{natbib} 
\usepackage{mathtools}  
\usepackage{titlesec}  
\usepackage{lastpage}  
\usepackage{tabularx}
\usepackage{booktabs}
\usepackage[english]{babel}  
\usepackage{amsmath}  
\usepackage{wasysym}  
\usepackage{bbm}  
\usepackage{array}  
\usepackage{xr}  
\usepackage{verbatim} 
\usepackage{float} 
\usepackage{xcolor}
\usepackage{multirow}
\usepackage{enumitem}

\usepackage{amsmath,amssymb,amsfonts}
\usepackage{algorithm}
\usepackage{algpseudocode}
\usepackage{graphicx}
\usepackage{textcomp}

\newcommand{\norm}[1]{\left\lVert#1\right\rVert}

\definecolor{blue1}{RGB}{79, 113, 190}
\definecolor{orange1}{RGB}{239, 137, 51}
\definecolor{graynew}{gray}{0.88}
\definecolor{purple1}{RGB}{112, 48, 160}

\hypersetup{%
  pdfborderstyle={/S/U/W 1}
}

\setcitestyle{round,numbers}

\newcommand{\email}[1]{\href{mailto:{#1}}{{#1}}}

\newcommand{\keywords}[1]{\textbf{Keywords}: {#1}}

\newcommand{\optincludegraphics}[2][]{}

\newcommand{\optinput}[1]{}

\newtagform{brackets}{[}{]}
\usetagform{brackets}

\newcommand{\thejournal}[1]{Magnetic Resonance in Medicine}

\title{Accelerating Stroke MRI with Diffusion Probabilistic Models through Large-Scale Pre-training and Target-Specific Fine-Tuning}

\immediate\write18{texcount -sum=1,1,1,1,1,1,1 -merge -incbib -dir -utf8 \jobname.tex > \jobname.wcTotal }
\immediate\write18{texcount -sub=none -merge -incbib -dir -utf8 \jobname.tex | grep _main_ | grep -oE '[0-9]+' | python -c "import sys; print(sum(int(l) * w for l, w in zip(sys.stdin, [1, 1, 0, 0, 0, 0, 0])))" > \jobname.wcManuscript }
\immediate\write18{texcount -sub=none -merge -incbib -dir -utf8 \jobname.tex | grep Abstract | grep -oE '[0-9]+' | python -c "import sys; print(sum(int(l) * w for l, w in zip(sys.stdin, [1, 1, 0, 0, 0, 0, 0])))" > \jobname.wcAbstract }

\newcommand{\wcTotal}{\clearpage{\noindent\large{\bf Detailed Word Count} (not to be included for submission)}\verbatiminput{\jobname.wcTotal}}
\newcommand{\wcManuscript}{\input{\jobname.wcManuscript}}
\newcommand{\wcAbstract}{\input{\jobname.wcAbstract}}

\begin{document}

\begin{titlepage}
{\noindent\LARGE\bf \thetitle}

\bigskip

\begin{flushleft}\large
	Yamin Arefeen\textsuperscript{1,2,$\dagger$},
	Sidharth Kumar\textsuperscript{1,$\dagger$},
    Steven Warach\textsuperscript{3},
    Hamidreza Saber\textsuperscript{3,4},
    Jonathan I. Tamir\textsuperscript{1,5,6}
\end{flushleft}


\noindent
\textsuperscript{$\dagger$}These authors contributed equally to this work.
\begin{enumerate}[label=\textbf{\arabic*}]
\item{Chandra Family Department of Electrical and Computer Engineering\\The University of Texas at Austin\\Austin, TX, USA}
\item{Department of Imaging Physics\\MD Anderson Cancer Center\\Houston, TX, USA}
\item{Dell Medical School Department of Neurology\\The University of Texas at Austin\\Austin, TX, USA}
\item{Dell Medical School Department of Neurosurgery\\The University of Texas at Austin\\Austin, TX, USA}
\item{Dell Medical School Department of Diagnostic Medicine\\The University of Texas at Austin\\Austin, TX, USA}
\item{Oden Institute for Computational Engineering and Sciences\\The University of Texas at Austin\\Austin, TX, USA}
\end{enumerate}

\bigskip


\textbf{*} Corresponding author:

\indent\indent
\begin{tabular}{>{\bfseries}rl}
Name		& Yamin Arefeen
\\
Department	& Chandra Family Department of Electrical Engineering and Computer Science \\
Institute	& The University of Texas at Austin; Austin, TX, United States\\
E-mail		& \email{yaminarefeen@gmail.com}											\\
\end{tabular}\\

Word Count: $250$ (Abstract) $\sim5000$ (body)\\
\indent Figure/Table Count: 10
\end{titlepage}

\begin{abstract}
\noindent\textbf{Purpose:} To develop a data-efficient strategy for accelerated MRI reconstruction with Diffusion Probabilistic Generative Models (DPMs) that enables faster scan times in clinical stroke MRI when only limited fully-sampled data samples are available.\\
\noindent\textbf{Methods:} Our simple training strategy, inspired by the foundation model paradigm, first trains a DPM on a large, diverse collection of publicly available brain MRI data in fastMRI and then fine-tunes on a small dataset from the target application using carefully selected learning rates and fine-tuning durations. The approach is evaluated on controlled fastMRI experiments and on clinical stroke MRI data with a blinded clinical reader study.\\
\noindent\textbf{Results:} DPMs pre-trained on approximately 4000 subjects with non-FLAIR contrasts and fine-tuned on FLAIR data from only 20 target subjects achieve reconstruction performance comparable to models trained with substantially more target-domain FLAIR data across multiple acceleration factors. Experiments reveal that moderate fine-tuning with a reduced learning rate yields improved performance, while insufficient or excessive fine-tuning degrades reconstruction quality. When applied to clinical stroke MRI, a blinded reader study involving two neuroradiologists indicates that images reconstructed using the proposed approach from $2 \times$ accelerated data are non-inferior to standard-of-care in terms of image quality and structural delineation.\\
\noindent\textbf{Conclusion:} Large-scale pre-training combined with targeted fine-tuning enables DPM-based MRI reconstruction in data-constrained, accelerated clinical stroke MRI. The proposed approach substantially reduces the need for large application-specific datasets while maintaining clinically acceptable image quality, supporting the use of foundation-inspired diffusion models for accelerated MRI in targeted applications.\\
\noindent\keywords{Accelerated MRI; Diffusion probabilistic models;  Foundation models; Stroke MRI; Clinical Translation}
\end{abstract}
\newpage

\section{Introduction}
Foundation models \cite{bommasani2021opportunities} represent a powerful paradigm in machine learning where models pre-trained on large, diverse datasets adapt effectively to target applications in a zero-shot fashion or with fine-tuning \cite{zhou2024foundationsurvey}. As a non-exhaustive list, this approach successfully applies across many domains, including natural language processing \cite{zhao2023llmsurvey}, vision \cite{radford2021learning}, segmentation \cite{kirillov2023sam}, fluorescence microscopy \cite{ma2024foundmicro}, and various healthcare applications \cite{huang2023twitter,zhou2023retinal,moor2023genmedai}.

Machine learning currently achieves state-of-the-art performance in reconstructing an MRI image from under-sampled data for many anatomy, contrasts, and datasets \cite{heckel2024ml,muckley2021fastmri,lyu2023cmrchallenge,desai2024skm}. 
However, these models are typically trained on datasets specific to the application of interest and are applied without modification for inference.
While effective given sufficient data, this approach degrades in settings with limited data to train the models for the desired application \cite{heckel2024ml}. 

As a motivating example, medical imaging is critical to diagnosis of ischemic stroke in patients who come to the emergency room presenting symptoms; many sites use computed tomography (CT) due to its effectiveness, widespread availability, and relatively short scan times \cite{birenbaum2011imaging}. However, MRI offers improved sensitivity when detecting ischemic stroke and more precise localization of the ischemic lesion \cite{rapillo2024strokemr}. Despite these advantages, MRI typically requires longer scan times and is more susceptible to degradation from patient motion, which can delay image acquisition and, consequently, time to treatment. Machine learning–based acceleration methods could mitigate these limitations by reducing scan duration and motion sensitivity, but limited stroke specific training data precludes their application.

Several techniques have been proposed to train machine learning reconstruction models in settings with limited, fully-sampled data. Some methods train reconstruction models on just the under-sampled scan itself by leveraging calibration regions \cite{mehmet2019raki,yamin2022spark}, implicit regularization imposed by neural network structure \cite{feng2024imjense,darestani2021convdecoder}, and self-supervised learning through k-space splitting \cite{yaman2022ssduzs}. However, these methods fall short of state-of-the-art reconstruction quality offered by methods trained on large amounts of high quality data \cite{lyu2023cmrchallenge,muckley2021fastmri,heckel2024ml}. Other self-supervised training strategies instead utilize under-sampled or corrupt images for training \cite{yaman2020ssdu,millard2023ssdu,desai2023noise2recon,wang2025selfsupervisedbench,derastani2022test} but still require a large corpus of data. In addition, these methods implement end-to-end reconstructions mapping under-sampled data to images given a specific acquisition model.

Motivated by data limitations in our stroke application and inspired by the foundation model paradigm \cite{terris2026ram}, we propose a method to train diffusion probabilistic generative models (DPMs) for accelerated MRI reconstruction given a small dataset for the contrast of interest.
Inspired by similar approaches for large language models \cite{parmar2024reuse}, our simple but effective strategy first trains a DPM on a large, public dataset of diverse MR images and then fine-tunes the DPM on the smaller, fully-sampled dataset from the target application. 
During fine-tuning, we select hyperparameters to adapt the model effectively without overfitting and retaining the benefits of pre-training on a large, diverse dataset \cite{lin2024diversity}. 
We select DPMs for the reconstruction model because they are agnostic to the acquisition forward model \cite{song2021sde,chung2022score,chung2023dps,jalal2021robust,luo2023diff,yang2023diff,ho2020diff,zheng2025inversebench,peng2020bayesian,pruessmann2019bayesian}. 
Thus unlike end-to-end methods which require the larger external dataset and the smaller target dataset to be acquired with the same measurement model (sampling pattern, coil geometry, etc\dots) \cite{heckel2024ml}, a DPM pre-trained on the larger dataset can be adapted to smaller target datasets with different acquisition models. 

First, we comprehensively evaluate the proposed approach on fastMRI \cite{muckley2021fastmri} by initially pre-training a DPM on $T_1$, $T_2$, and $T_1$-post brain scans from 4000 subjects. To simulate access to limited data on a target application, we then finetune the DPM on FLAIR data from just 20 subjects of fastMRI. Across a range of acceleration factors, DPMs trained with our method achieve comparable performance to DPMs trained with 344 target FLAIR subjects. Our experiments also highlight best practices when selecting inference hyperparameters for DPM-based MRI reconstruction and showcase application on prospectively under-sampled data.

Next, we apply our method to the motivating setting of stroke MRI. We initially pre-train a DPM on $T_1$, $T_2$, $T_1$-post, and FLAIR brain scans from approximately 4000 fastMRI subjects and then fine-tune the DPM on data from 25 stroke patients from our local hospital. Testing on 5 additional patients demonstrates that the proposed method enables reconstruction of SWI, MPRAGE, DWI, and FLAIR images from $2\times$ further under-sampled stroke data.

Finally, we conduct a clinical validation study in which two radiologists experienced in stroke MRI compared SWI, MPRAGE, DWI, and FLAIR images reconstructed from retrospective, $2\times$ under-sampled data using our method to standard-of-care images. The reader study indicates that images reconstructed by the DPM pre-trained on fastMRI and fine-tuned on local stroke data achieves non-inferior image quality and structural delineation of anatomy relative to the standard-of-care.

\section{Methods}
\subsection{Diffusion Probabilistic Generative Models for MRI Reconstruction}
We model the multi-coil MRI acquisition as $y = P F S x + n$ where $y$ are the acquired measurements, $P$ is the under-sampling operator, $F$ is the Fourier Transform, $S$ is the multi-coil operator, $n$ is additive gaussian noise, and $x$ is the image to reconstruct. DPMs reconstruct accelerated MRI acquisitions by first learning the underlying probability distribution of the target MR contrast and anatomy, and then guiding the reconstruction towards solutions that both match the data and are statistically likely. Specifically, we use score-based diffusion models where a neural network $G_{\theta}$ approximates the score function $\nabla_{x_t} \log p_t(x)$, where $p_t(x)$ denotes image distribution perturbed by gaussian noise with standard deviation $\sigma_t$ \cite{song2021sde,karras2022edm,ho2020diff,yang2023diff}. Equipped with the score function approximation, images can be reconstructed from under-sampled data by approximately sampling from the posterior $p(x|y)$ with posterior sampling algorithms \cite{zheng2025inversebench}.

In this work, we combine the prior sampling formulation from Ref. \cite{karras2022edm} with Diffusion Posterior Sampling \cite{chung2023dps} to approximately sample from $p(x|y)$ in accelerated MRI. Specifically, we solve the following ordinary differential equation (ODE): 
\begin{equation}
    d\mathbf{x} = \left[\frac{\dot{s}(t)}{s(t)}\mathbf{x} - s(t)^2 \dot{\sigma}(t)\sigma(t)\nabla_{\mathbf{x}} \log p\left( \frac{\mathbf{x}}{s(t)} | \mathbf{y}; \sigma(t) \right)\right] dt.
\end{equation}
Setting $s(t)=1$ and $\sigma(t)=t$ based on Ref. \cite{karras2022edm}, applying Baye's rule to separate the posterior score into the sum of the log-likelihood and prior score, and inserting the DPM that approximates the prior score function and the closed-form expression for the likelihood score \cite{chung2023dps,jalal2021robust} results in the following ODE,
\begin{equation}
    d\mathbf{x} = \left[-t \left(\nabla_\mathbf{x} \norm{\mathbf{P F S}\mathbf{\tilde{x}}(\mathbf{x}) - \mathbf{y}}_2^2 + D_\theta(\mathbf{x},t\right) \right] dt.
    \label{eq:ode}
\end{equation} We apply the approximation $\mathbf{\tilde{x}}(\mathbf{x}) = \mathbb{E}[\mathbf{x_0}|\mathbf{x}]$ since the log-likelihood score is only known at $t=0$ \cite{chung2023dps}.

\subsection{DPM Details}
All DPMs in this work use the U-Net style architecture from Ref. \cite{song2021sde}, as implemented by Ref. \cite{karras2022edm}, and we adopt a modified version of the architecture that supports inputs with varying matrix sizes \cite{arefeen2025neonatal}. Initial training on the external dataset and fine-tuning on the target dataset applied the ``EDM'' training loss, data augmentation, optimizer choice, and noise level scheduling. 

To accommodate datasets with heterogeneous contrasts, we simultaneously train on all available data by conditioning the model on contrast type \cite{arefeen2025neonatal}. Each image contrast type in the training set is assigned a one-hot vector, which is passed through a small fully-connected neural network to produce an embedding. Each U-Net block in the DPM network then takes this embedding vector as an input, allowing the model to adapt its behaviour based on the contrast. The DPM and embedding network are trained jointly.

\subsection{Posterior Sampling Details}
Algorithm \ref{alg:DPS} outlines the posterior sampling procedure utilized in this work for accelerated MRI reconstruction adapted from the ODE prior sampling procedure used in Ref. \cite{karras2022edm} and the log-likelihood approximation proposed by Ref. \cite{chung2023dps}. The noise levels, $\hat{\sigma_i}$ follow the recommended schedule from Ref. \cite{karras2022edm}, and $N$ denotes the number of sampling steps. Since this posterior sampling is approximate, $\zeta_i$ is introduced as a tunable hyperparameter that balances data consistency and learned prior score at each iteration.

\begin{algorithm}[H]
\caption{Diffusion Posterior Sampling  (DPS)}\label{alg:DPS}
\begin{algorithmic}
\Require $G_{\theta}(x;\sigma), N, \{\zeta_{i=0}^{N-1}\}, \{\hat{\sigma}_i\}_{i=0}^{N-1}$ \\
\textbf{sample} $x_{t_0}\sim \mathbb{N}(0,I)$
    \For{$i \in \{0,...,N-1\}$} 
        \State $\hat{s} = \frac{G_{\theta}(x;\sigma) - x_i}{\sigma} $ 
        \State $\hat{x}_0 = \frac{1}{\sqrt{\alpha_i}} (x_i + (1-\bar{\alpha}_i\hat{s}) $ 
        \State $n\sim\mathbb{N}(0,I)$
        \State $\tilde{x}_p = \frac{\sqrt{\alpha_i}(1-\bar{\alpha}_{i-1})}{1-\bar{\alpha}_i}x_i + \frac{\sqrt{\bar{\alpha}_{i-1}}\beta_i}{1-\bar{\alpha}_{i-1}}\hat{x}_0$
        \State $\tilde{x}_{DC} = \nabla_{x_i}||y - A(\hat{x}_0)||^2_2$
        \State $x_{i+1} = \tilde{x}_p - \zeta_i \tilde{x}_{DC} +\sigma_i n$
    \EndFor
\end{algorithmic}
\end{algorithm}

\subsection{Proposed Training Strategy}
\label{sec:proposed}
\begin{figure}[H]
    \centering
    \centerline{\includegraphics[width=\linewidth]{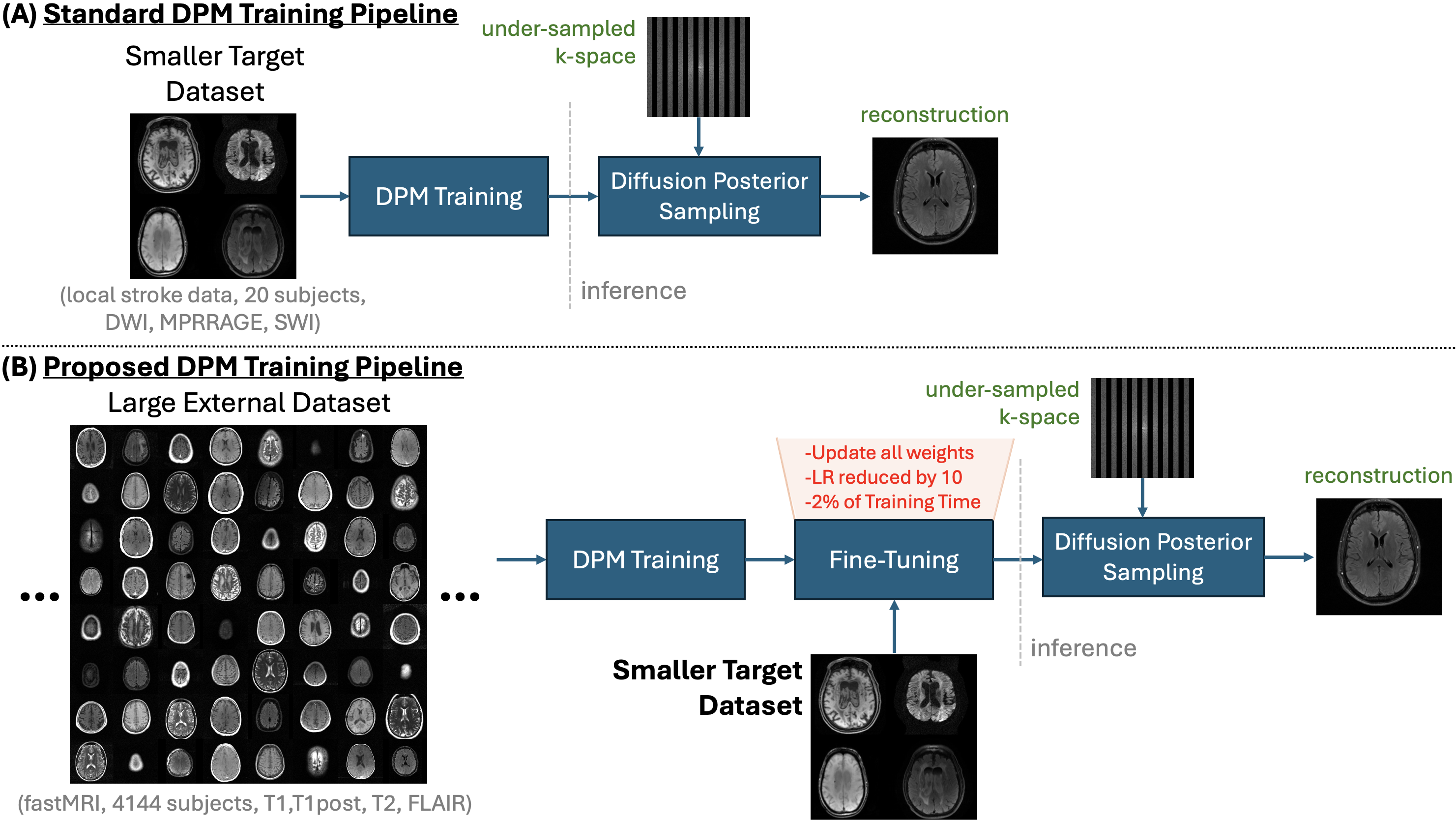}}
    \caption{(A) The standard DPM training pipeline trains exclusively on the target-domain dataset, but if the target dataset is small, this can lead to degraded performance. (B) The proposed training pipeline mitigates data scarcity by first pre-training the DPM on a large, diverse external MRI dataset, followed by fine-tuning on a small target-domain dataset using a reduced learning rate and limited number of epochs. The fine-tuned model is then used for posterior sampling–based reconstruction from under-sampled k-space.}
    \label{fig:schematic}
\end{figure}

Inspired by strategies for foundation and large language models \cite{parmar2024reuse}, we propose a simple and effective training approach to train DPMs for accelerated MRI when only limited target-domain data is available. Our strategy involves two stages: (i) pre-training on a large, diverse external dataset, like fastMRI \cite{muckley2021fastmri}, covering multiple MRI contrasts; and (ii) fine-tuning on the smaller, target dataset, updating all model weights. To mitigate overfitting during fine-tuning, we reduce the learning rate by an order of magnitude and train for a small number of epochs, corresponding to roughly 2\% of pre-training time. In addition, we assign the target contrast its own one-hot encoded vector and update the weights of the embedding network during fine-tuning. Figure \ref{fig:schematic} compares (A) standard and (B) the proposed training strategies for DPMs.

\subsection{Controlled Experiments with FastMRI}
To validate the proposed training strategy, we use the multi-coil brain dataset from FastMRI \cite{muckley2021fastmri}. Our dataset included Axial $T_2$ from 2678 subjects, Axial $T_1$ from 498 subjects, post-contrast Axial $T_1$ from 949 subjects, and Axial FLAIR from 344 subjects. We focused on reconstructing Axial FLAIR images, simulating a low-data regime by limiting the FLAIR training set to only 20 subjects. All k-space data were pre-whitened and re-sized to $320 \times 320$ matrix size, and training was performed on fully-sampled, coil-combined images normalized to the 99th percentile of the maximum absolute value. For testing, 500 random Axial FLAIR slices were selected from held-out subjects not used during training.

We first investigated the effect of learning rate and training time during fine-tuning. A DPM was pre-trained on the larger external dataset with a learning rate of $10^{-4}$ for $625,000$ epochs, and then fine-tuned on the 20-subject FLAIR dataset with learning rates of $\{5\times 10^{-4}, 1\times 10^{-4}, 5\times 10^{-5}, 5\times 10^{-5}\}$ and epochs of $\{0, 625, 1250, 1875, 2500\}$.

In addition, we compared our proposed training paradigm to alternative methods. 
\begin{itemize}
    \item \textbf{Method 1:} Trained a DPM on the full dataset, using all 2678 Axial $T_2$,  498 Axial $T_1$, 949 Post-contrast Axial $T_1$, and 344 Axial FLAIR subjects.
    \item \textbf{Method 2:} Pre-trained on $T_1$, $T_2$ and Post-contrast $T_1$ data and fine-tuned on all 344 FLAIR subjects.
    \item \textbf{Method 3:} Trained only on the 344-subject FLAIR dataset.
    \item \textbf{Method 4 (Proposed):} Pre-trained on all $T_2$, $T_1$, and $T_1$ post-contrast data and then fine-tuned on 20 FLAIR subjects with reduced learning rate and shorter training time.
    \item \textbf{Method 5:} Trained solely on the 20-subject FLAIR dataset.
    \item \textbf{Method 6:} Trained jointly on the full external dataset and the 20-subject FLAIR dataset without fine-tuning.
\end{itemize}  
Methods 1-3 serve as performance upper bounds, as they use extensive target contrast (FLAIR) data. Methods 4-6 represent the constrained data regime, relying on large external datasets and limited target-domain data.

All methods were evaluated on the 500 validation FLAIR slices using under-sampling rates of $R = \{3,4,5,6,7,8\}$ with no auto-calibration (ACS) data.

\subsection{Evaluating Posterior Sampling Hyperparameters}
We evaluated the effect of hyperparameter selection in Algorithm \ref{alg:DPS} for approximate posterior sampling. In particular, the data consistency weighting $\zeta$ is often set to be inversely proportional to the norm of the data consistency error \cite{chung2023dps}, while the number of steps $N$ trades off between quality and computation. Given a DPM trained for FLAIR reconstruction with the proposed method, we compared reconstruction performance for different choices of $\zeta$ between $1/2$ and $5$ and $N$ between $200$ and $300$ when reconstructing data at $R = \{3,4,5,6,7,8\}$.

\subsection{Application to Clinical Stroke MRI}
In collaboration with our local hospital's medical school, under IRB approval and informed consent, we collected a small dataset of clinical stroke MRI from 30 patients on a 3T Siemens Vida scanner. Each subject underwent a standard-of-care stroke imaging protocol following triage, including T2-FLAIR, MPRAGE, SWI, and DWI. The standard-of-care data were reconstructed with SENSE-based parallel imaging using coil maps calibrated with BART \cite{bart}. The list below summarizes the acquisition parameters:
\begin{itemize}
    \item T2-FLAIR: $0.5 \times 0.5 \text{ mm}$ in-plane resolution, $5 \text{ mm }$ slice thickness, $9000 \text{ ms}$ TR.
    \item MPRAGE: $1.5 \times 1.5 \times 1.5 \text{ mm}$ resolution, $240 \times 240 \times 168 \text{ mm}$ FOV, $2500 \text{ ms}$ TR.
    \item SWI: $1.0 \times 1.0 \text{ mm}$ resolution, $250 \times 211 \text{ mm}$ FOV, $2.5 \text{ mm}$ slice thickness.
    \item  DWI: $0.8 \times 0.8 \text{ mm}$ resolution, $250 \times 250 \text{ mm}$ FOV, 30 slices, $5.0 \text{ mm}$ slice thickness, b-values: $0, 1200 \text{ s/mm}^2$.
\end{itemize}

To evaluate our method in this clinical setting, we first pre-trained a DPM on all Axial FLAIR, Axial $T_2$, Axial $T_1$, and Axial $T_1$ post contrast data available from the 4125 subjects in fastMRI. Then, we fine-tuned the DPM seperately for each clinical sequence (T2-FLAIR, MPRAGE, SWI, DWI) using 20 stroke patients from our local dataset. Fine-tuning was performed with a reduced learning rate and limited training duration as described in Section \ref{sec:proposed}. 

An additional 5 stroke subjects were used for validation to select hyper-parameters for fine-tuning and posterior sampling. Fine-tuned DPMs then reconstructed retrospectively under-sampled data (accelerations around $R \approx 4$) on the remaining 5 test subjects. 

\subsection{Clinical Stroke Reader Study}
We acquired additional clinical stroke MRI data from 80 subjects at our local hospital with IRB approval and informed consent to conduct a blinded reader study. Each subject was scanned using the standard-of-care stroke protocol, including T2-FLAIR, MPRAGE, SWI, and DWI acquisitions obtained at an acceleration factor of approximately $R \approx 2$ and reconstructed using SENSE-based parallel imaging.  To evaluate accelerated imaging, the same data were retrospectively under-sampled by an additional factor of two, resulting in a total effective acceleration of approximately $R \approx 4$, and reconstructed using diffusion probabilistic models (DPMs) fine-tuned with the proposed training strategy.

For each of the 80 subjects, the reconstructed images corresponding to the standard-of-care and proposed accelerated protocols were grouped by subject, blinded, and randomized, yielding a total of 160 image sets for evaluation.  Importantly, each image set comprised contrasts from the stroke protocol, and readers assessed the image set as a whole rather than individual sequences in isolation. Reader H.S. evaluated all 160 image sets, while reader S.W. evaluated 42 image sets (corresponding to 21 subjects) due to time constraints; H.S. and S.W. have 11 and 30 years of clinical experience, respectively.

Each image set was scored for structural delineation of white matter, gray matter, and ventricles, as well as for image quality metrics including signal-to-noise ratio (SNR), contrast, sharpness, artifacts, and overall image quality. All evaluations were performed using a five-point ordinal scale where $[1 \leftarrow \text{repeat scan needed}, 2 \leftarrow \text{limited}, 3 \leftarrow \text{diagnostic}, 4 \leftarrow \text{good}, 5 \leftarrow \text{outstanding}]$. Statistically significant differences in metrics was assessed using paired two-sided Wilcoxon signed-rank tests and inter-rater reliability was measured with the Cohen's kappa coefficient.

\subsection{Experiments on Prospectively Under-sampled Acquisitions}
Noting that our clinical evaluation involved retrospective subsampling, we further evaluated our method with prospective under-sampling. We acquired SWI and FLAIR data from a healthy subject under IRB approval and informed consent using the parameters of the typical stroke protocol at our local hospital. Two acquisitions were performed: one at the standard-of-care acceleration rate, $R = 1.7$, and another at a prospective acceleration rate of $R = 3.75$. We applied the proposed method, pre-trained on all $T_2$, $T_1$, and post-contrast $T_1$ from fastMRI and fine-tuned on our FLAIR and SWI stroke data, to reconstruct the data prospectively  under-sampled to $R=3.41$ and $R=3.75$ respectively and to reconstruct the $R=1.75$ under-sampled data after it was further retrospectively under-sampled by $R=3.52$ and $R=3.75$ respectively. Reconstructions on both datasets were compared to a L1-wavelet baseline implemented with BART \cite{bart}.

\section{Results}
\subsection{Controlled Experiments with FastMRI}
\begin{figure}[t]
    \centering
    \centerline{\includegraphics[width=\linewidth]{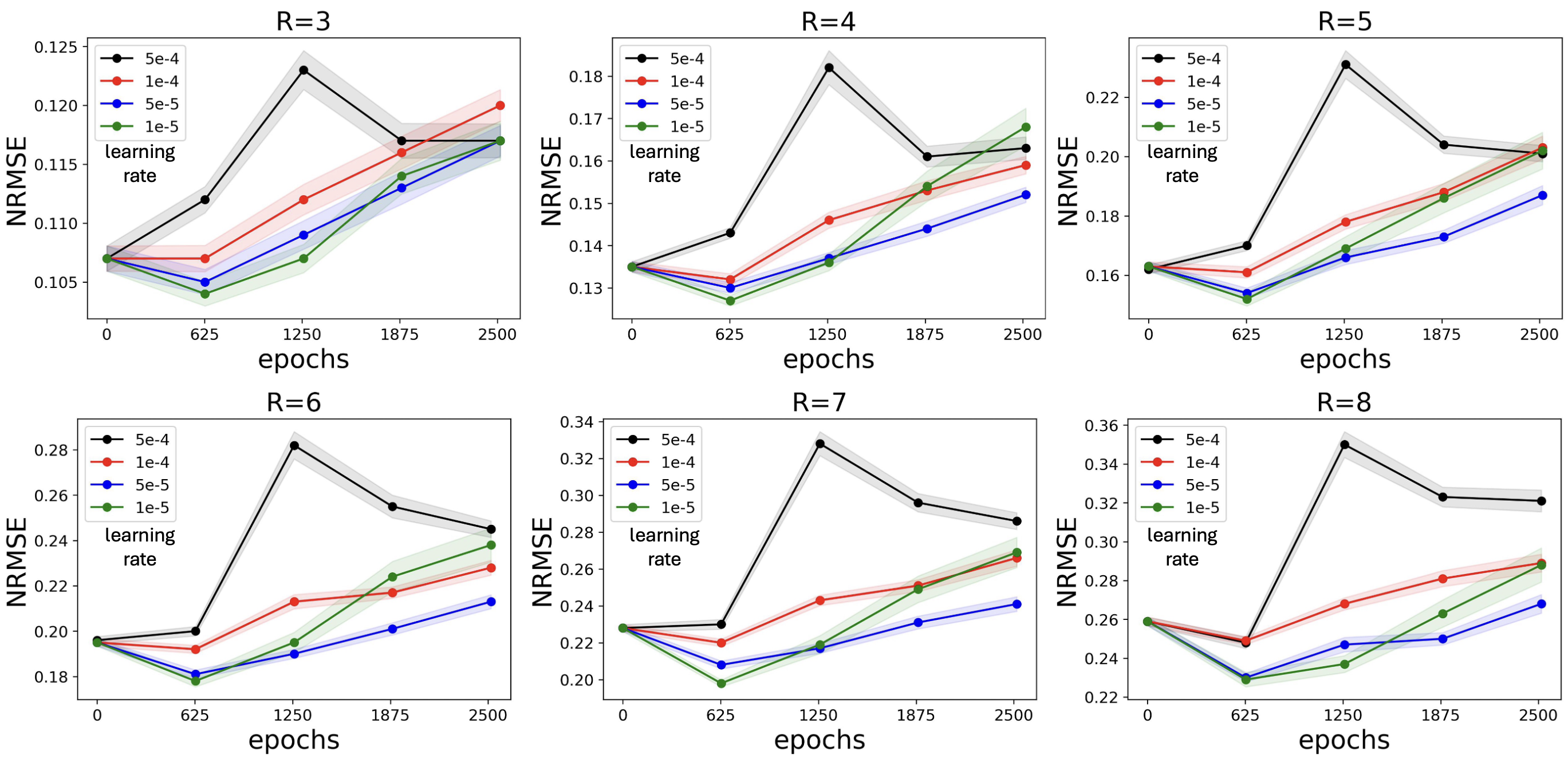}}
    \caption{Quantitative results from the controlled fastMRI experiments where DPMs were fine-tuned on 20 subjects to reconstruct fastMRI flair data. NRMSE on 500 validation slices as a function of fine-tuning duration (epochs) is plotted for different learning rates and acceleration factors.  Across all acceleration factors, moderate fine-tuning improves reconstruction quality relative to no fine-tuning, while excessive fine-tuning leads to degraded performance, particularly at higher learning rates.}
    \label{fig:flaireval_quant}
\end{figure}

Figure \ref{fig:flaireval_quant} quantitatively evaluates the proposed fine-tuning method across different learning rates and training epochs, measured with normalized root mean squared error (NRMSE) at various acceleration factors on the 500 fastMRI FLAIR validation slices. Across all acceleration rates and learning rates, fine-tuning for 650 epochs yields the lowest NRMSE, outperforming both no fine-tuning (0 epochs) and longer (1250, 1875, 2500 epochs) fine-tuning durations. Similarly, a learning rate of $1\times10^{-5}$ achieves the best performance in comparison to higher learning rates.

\begin{figure}[t]
    \centering
    \centerline{\includegraphics[width=\linewidth]{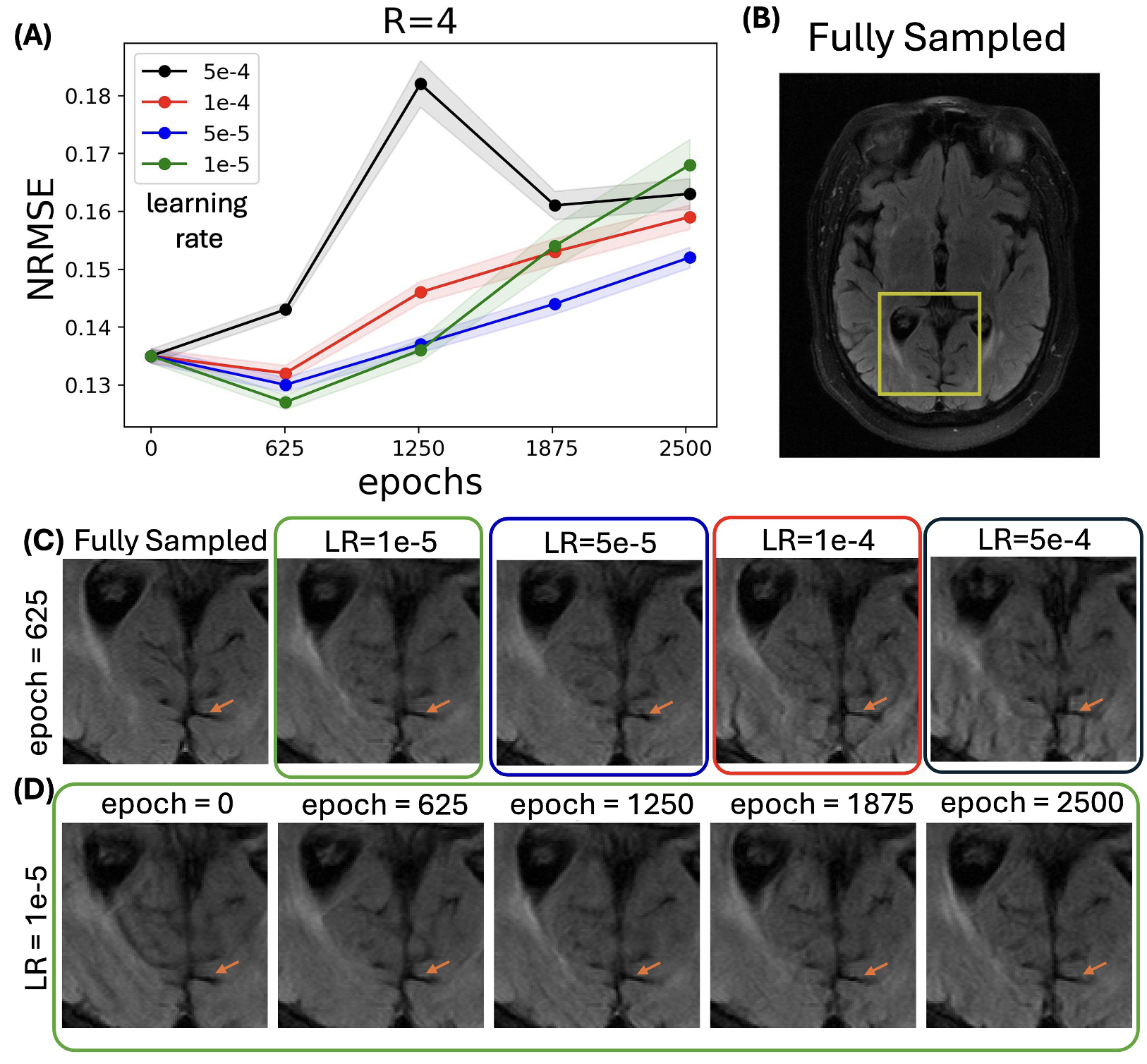}}
    \caption{A representative example from the fastMRI FLAIR experiment. (A) Quantitative NRMSE as a function of fine-tuning duration for different learning rates, highlighting that a comparatively smaller learning rate and fewer fine-tuning epochs yields the best performance. (B) Fully sampled reference FLAIR image. (C) Zoomed reconstructions obtained using different learning rates with 650 fine-tuning epochs, compared against the fully sampled reference. (D) Reconstructions obtained using a fixed learning rate of $1\times10^{-5}$ across different fine-tuning durations. Appropriate fine-tuning hyperparameters lead to visibly improved structural fidelity and reduced artifacts, as highlighted by the orange arrows.}
    \label{fig:flaireval_qual}
\end{figure}

Figure \ref{fig:flaireval_qual} illustrates a representative example from the quantitative FLAIR experiment at $R = 4$. As shown in (A), combining a learning rate of $1\times10^{-5}$ and 650 epochs gives the best quantitative performance. (B) shows the fully-sampled reference and (C) zooms into the reference and reconstructions under various learning rates and epochs. Qualitative improvements are observed, highlighted by the orange arrow, when the appropriate fine-tuning hyperparameters are used. 

\begin{figure}[t]
    \centering
    \centerline{\includegraphics[width=\linewidth]{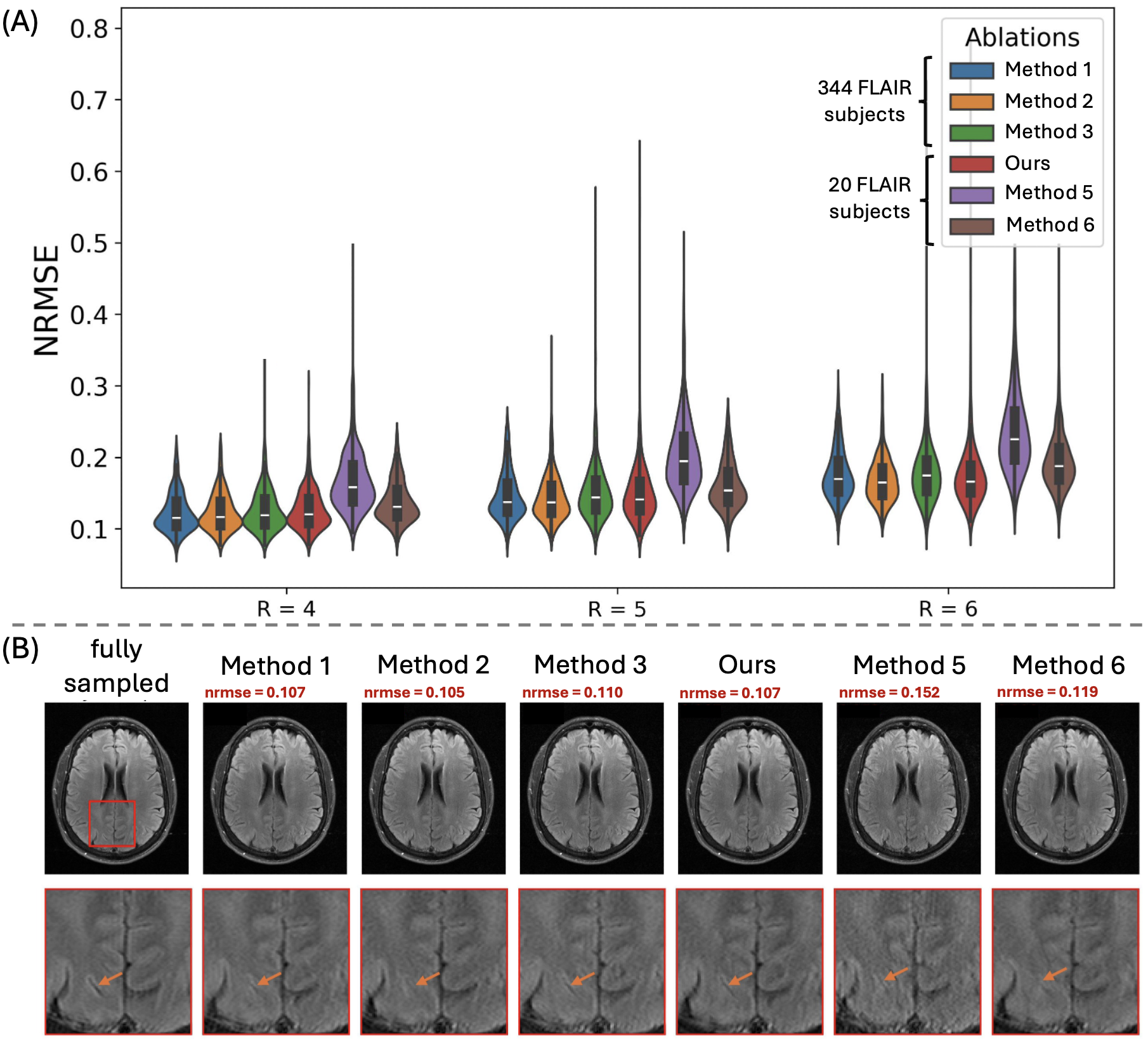}}
    \caption{(A) DPMs are compared across methods with access to different amounts of target-domain FLAIR data. Methods 1–3 are trained using FLAIR data from all 344 subjects, while Methods 5 and 6 are trained using 20 FLAIR subjects and the external dataset. The proposed approach (Method 4), pre-trained on a large external dataset and fine-tuned on 20 FLAIR subjects, achieves comparable NRMSE to Methods 1-3, despite using substantially less target domain data and consistently outperforms Methods 5,6 trained on the same limited data. (B) displays example reconstructions from this experiment.}
    \label{fig:method_comparisons}
\end{figure}

Figure \ref{fig:method_comparisons} compares DPMs fine-tuned using our method (650 epochs and $1\times10^{-5}$ learning rate) on FLAIR data from 20 subjects to methods (1-3) which have access to FLAIR data from 344 subjects and methods (5,6) with access to the same 20 FLAIR subjects and large external dataset. (A) plots quantitative reconstruction performance for acceleration rates $R=\{4,5,6\}$. Our method achieves comparable performance to methods 1-3 which have access to much more FLAIR data and improved performance in comparison to methods 5,6 with access to similar amounts of data. (B) shows example reconstructions from $R=4$ under-sampled data. Quantitatively and qualitatively, as highlighted by the orange arrow, our method achieves comparable quality to methods 1-3 and outperforms methods 5 and 6.

\begin{figure}[t]
    \centering
    \centerline{\includegraphics[width=\linewidth]{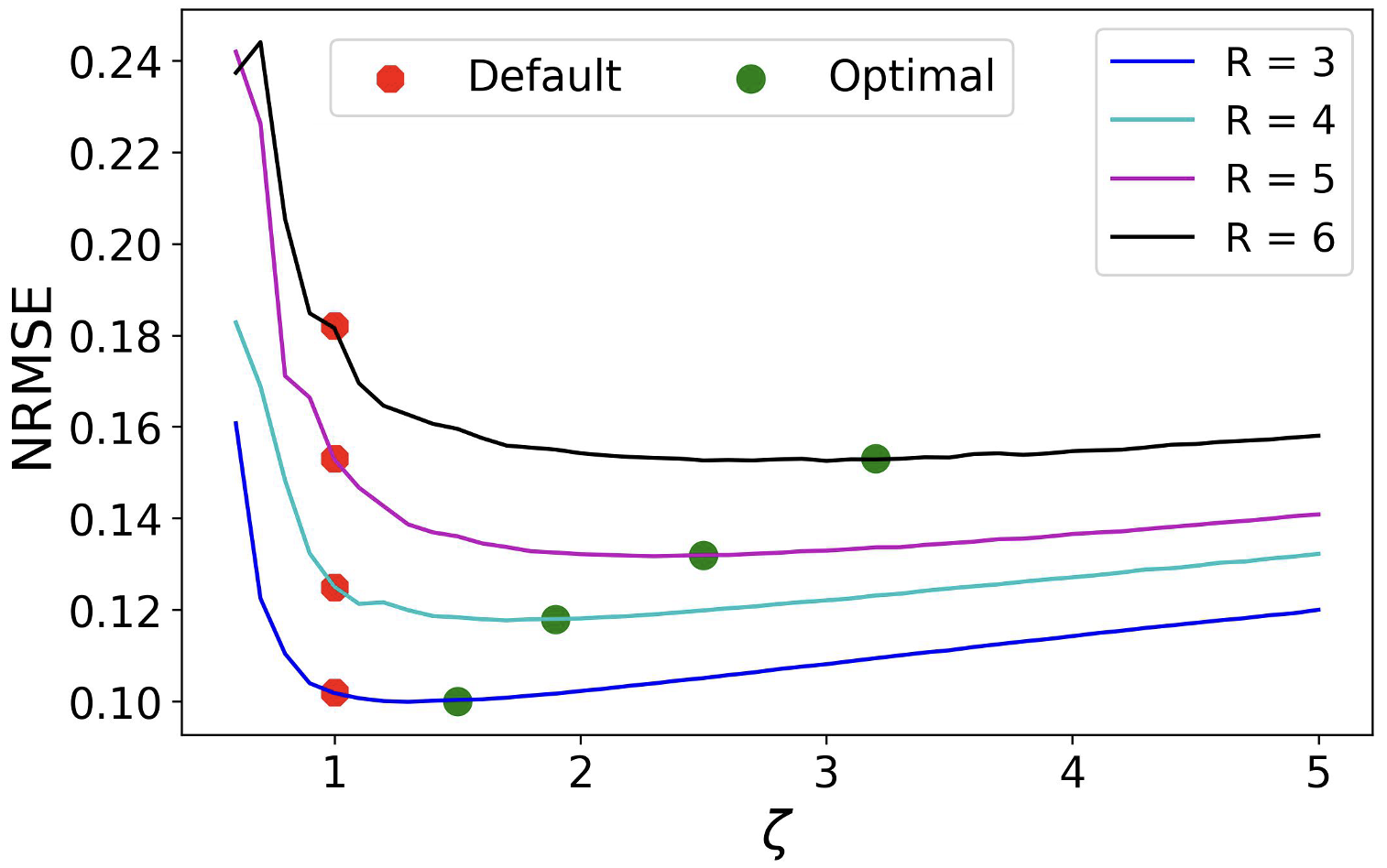}}
    \caption{Using the best fine-tuned DPM, average NRMSE on fastMRI FLAIR validation data is plotted as a function of the data-consistency step size $\zeta$ for multiple acceleration factors. Red markers denote the default choice $\zeta = 1$, while green markers indicate the value of $\zeta$ that minimizes reconstruction error at each acceleration. The optimal $\zeta$ increases with acceleration.}
    \label{fig:dps_hyper}
\end{figure}

Using our best fine-tuned DPM, Figure \ref{fig:dps_hyper} plots the average NRMSE across the validation samples for varying acceleration rates as a function of the data consistency step-size $\zeta$ in Algorithm \ref{alg:DPS}. Green dots highlight the typical default setting of $\zeta=1$ and the red dots mark the value of $\zeta$ that achieves the lowest reconstruction error at each acceleration rate. The optimal choice of $\zeta$ increases with higher acceleration rate. Supporting Table S1 summarizes the best performing of $\zeta$ and number of posterior sampling steps $N$ for each acceleration rate.

\subsection{Application to Clinical Stroke MRI}
\begin{figure}[t]
    \centering
    \centerline{\includegraphics[width=\linewidth]{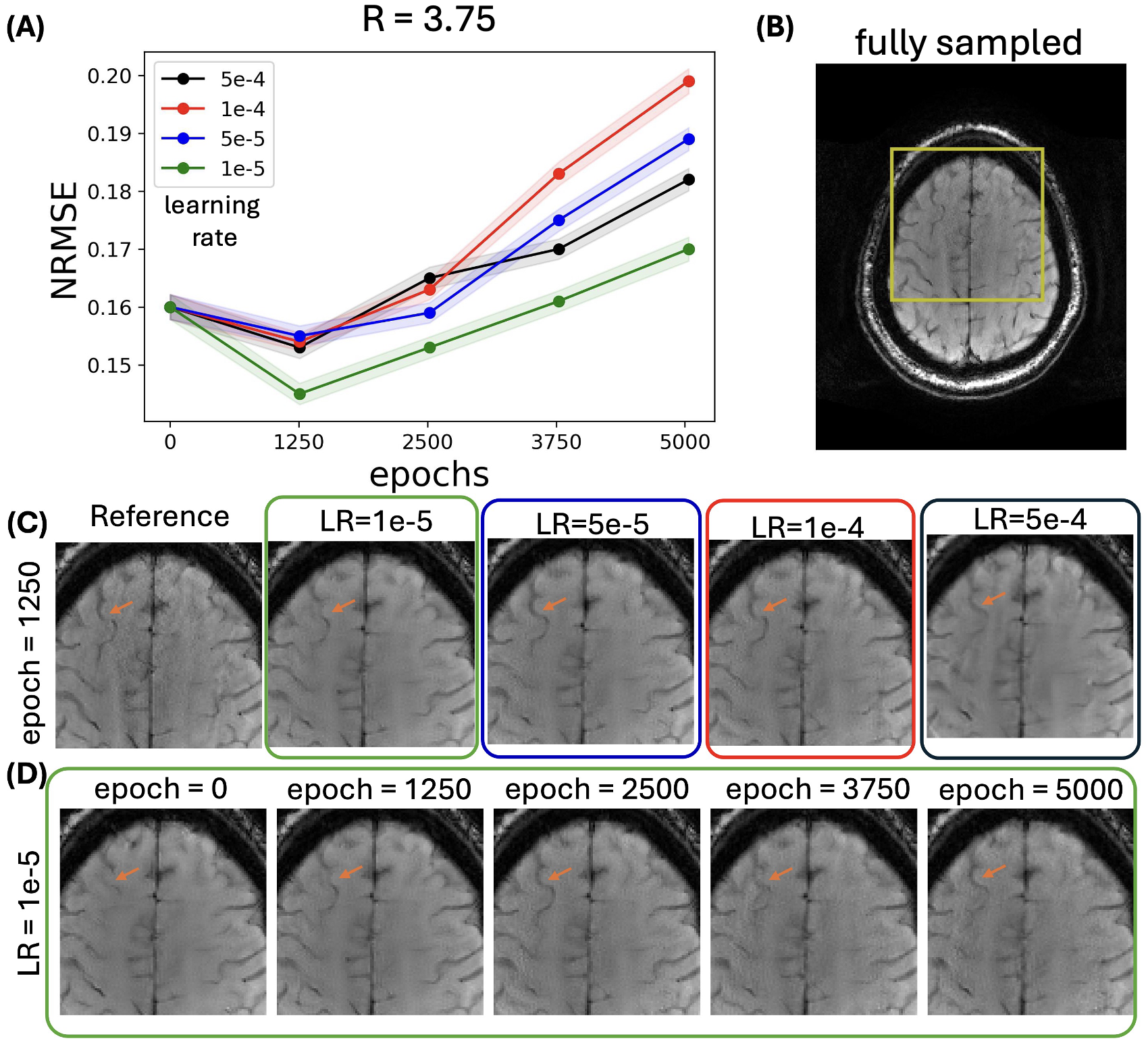}}
    \caption{(A) NRMSE of SWI reconstructions averaged over five validation subjects from data retrospectively accelerated by $R\approx 3.75$. (B) Fully-sampled reference. (C) Zoomed-in reconstructions obtained with fixed epochs and varying learning rate and (D) fixed learning rate and varying epochs. Similar to the fastMRI FLAIR experiments, insufficient or excessive fine-tuning duration and higher learning rates consistently result in worse reconstructions, demonstrating that the proposed training strategy is effective for clinical stroke SWI data.}
    \label{fig:swi_example}
\end{figure}

Figure \ref{fig:swi_example} evaluates DPMs fine-tuned using the proposed technique with limited stroke data to reconstruct $R \approx 3.75$ SWI data from clinical stroke acquisitions. Similar to FLAIR in the fastMRI setting, the NRMSE plots in (A) averaged over the 5 validation subjects suggest that fine-tuning for too few or too many epochs results in worse quantitative performance. In addition, higher learning rates yield worse performance for all training epochs. These results suggest that 1250 epochs and a learning rate of $1 \times 10^{-5}$ yields improved performance for fine-tuning a DPM to stroke SWI. (B) shows an example standard-of-care SWI image and (C) compares zoomed in views of this reference with reconstruction using DPMs fine-tuned with different hyperparameters. Analysis on the 5 subject validation dataset found $\zeta = 2.1$ as the optimal data consistency weighting and was used for all test reconstructions.

\begin{figure}[t]
    \centering
    \centerline{\includegraphics[width=\linewidth]{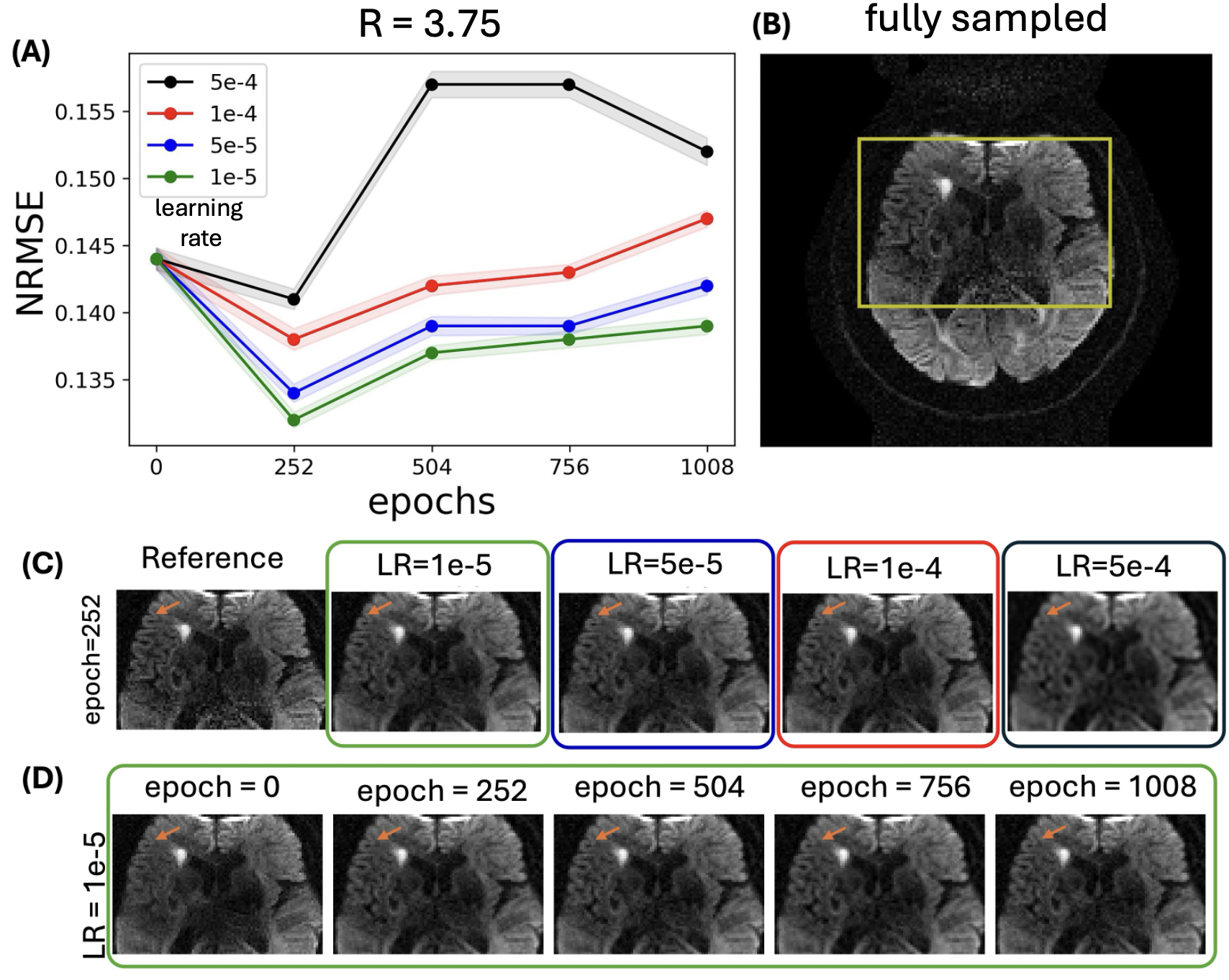}}
    \caption{(A) NRMSE of DWI reconstructions averaged over five validation subjects from data retrospectively accelerated by $R\approx 3.75$. (B) Fully-sampled reference. (C) Zoomed-in reconstructions obtained with fixed epochs and varying learning rate and (D) fixed learning rate and varying epochs. Similar to the fastMRI FLAIR experiments, insufficient or excessive fine-tuning duration and higher learning rates consistently result in worse reconstructions, demonstrating that the proposed training strategy is effective for clinical stroke DWI data.}
    \label{fig:dwi_example}
\end{figure}

Figure \ref{fig:dwi_example} presents fine-tuning a DPM to reconstruct DWI acquired with b-value $1200 \text{ s/mm}^2$. (A) Fine-tuning with 252 epochs and a learning rate of $1 \times 10^{-5}$ yields the best performance. (B) shows an example standard-of-care DWI image and (C) compares zoomed in views of the reference and reconstructions using our fine-tuned DPM with various hyperparameters.

Supporting Figure S1 analyzes the influence of fine-tuning epochs and learning rate for the additional FLAIR, MPRAGE, and DWI (b-value = 0) acquisitions in the test stroke data. Similar to SWI and DWI b-value $1200 \text{ s/mm}^2$, appropriate selection of fine-tuning hyperparameters improves reconstruction performance on $R\approx4$ under-sampled data. Supporting Figure S2 presents an example MPRAGE slice from the quantitative analysis of Supporting Figure S1.

\begin{figure}[t]
    \centering
    \centerline{\includegraphics[width=\linewidth]{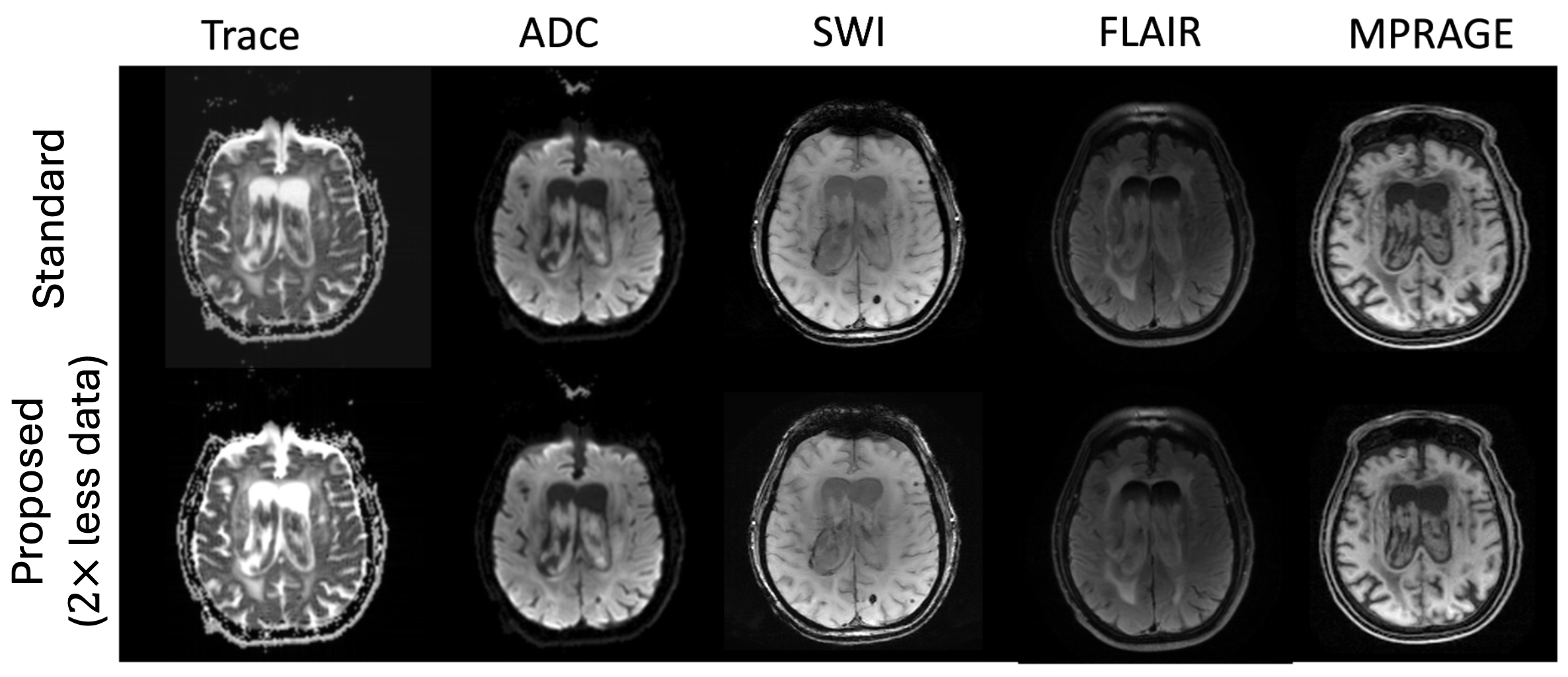}}
    \caption{Representative images from a clinical stroke protocol are shown for a single subject, comparing standard-of-care reconstructions (top row) with images retrospectively under-sampled by an additional factor of two and reconstructed using the proposed DPM-based approach (bottom row). The protocol includes diffusion-weighted imaging (shown as derived trace and apparent diffusion coefficient (ADC) maps), susceptibility-weighted imaging (SWI), FLAIR, and MPRAGE. Across all contrasts, the proposed accelerated reconstructions preserve structural detail and contrast comparable to the standard-of-care.}
    \label{fig:protocol_comparison}
\end{figure}

Figure \ref{fig:protocol_comparison} presents example stroke protocol images from the standard-of-care and images from the same protocol retrospectively under-sampled by an additional factor of 2 and reconstructed with DPMs fine-tuned with the proposed strategy. For DWI, we display derived trace and apparent diffusion coefficient (ADC) maps, commonly used in clinical stroke assessment, instead of displaying the raw DWI images directly.

\subsection{Clinical Reader Study}
\begin{figure}[t]
    \centering
    \centerline{\includegraphics[width=\linewidth]{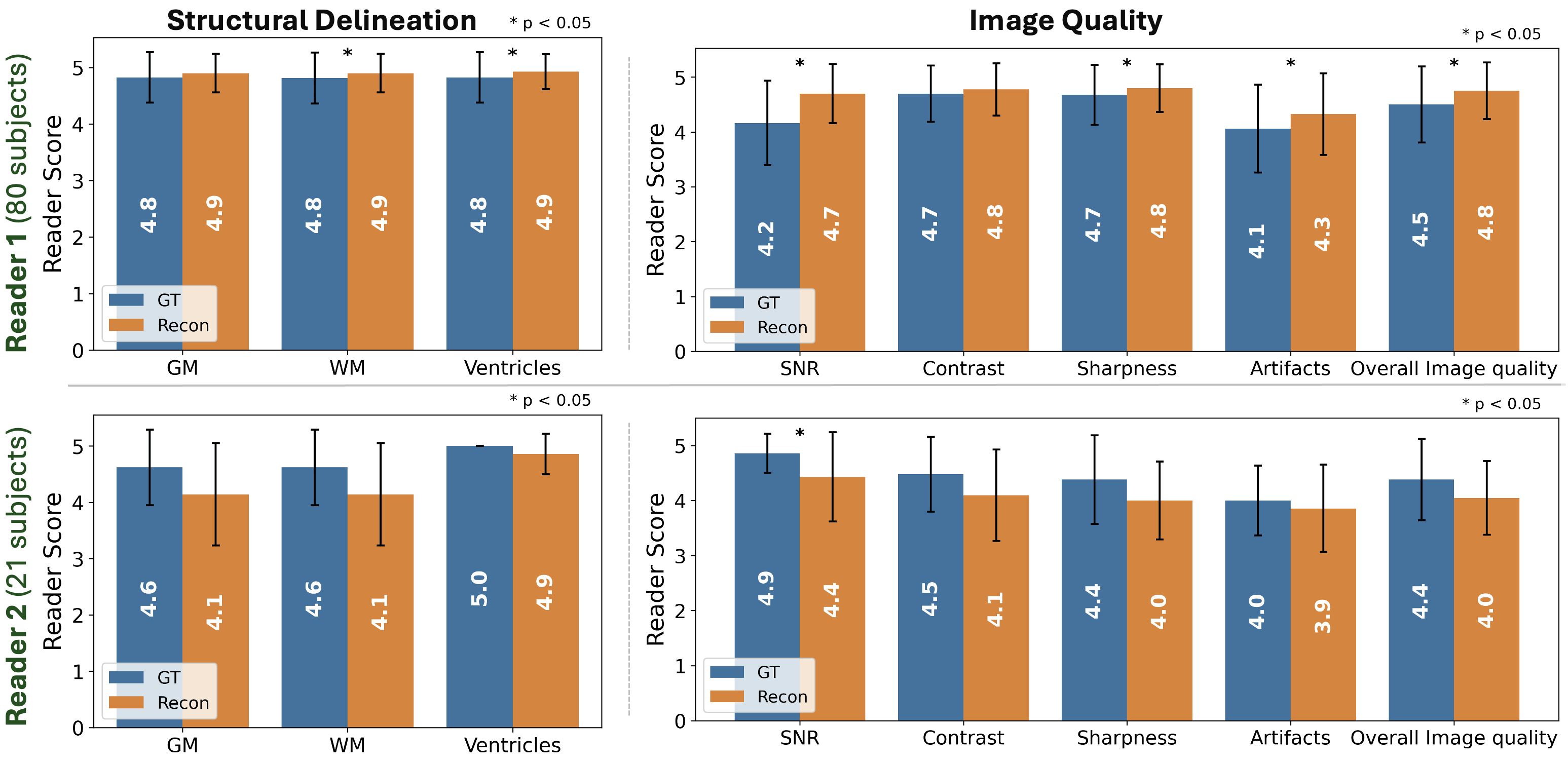}}
    \caption{Mean reader scores  for structural delineation (left) and image quality metrics (right) are shown for two board-certified neuroradiologists evaluating full stroke-protocol image sets reconstructed from standard-of-care data (GT) and from retrospectively accelerated data reconstructed using the proposed DPM-based approach (Recon). The top row reports results from Reader 1, who evaluated all 80 subjects, while the bottom row reports results from Reader 2, who evaluated 21 subjects. For Reader 1, the proposed accelerated reconstructions achieved statistically significant higher scores for several image quality metrics and for white matter and ventricular delineation. For Reader 2, no significant differences were observed except for SNR, where the standard-of-care reconstruction scored higher; however, mean scores for both methods remained high across all categories.}
    \label{fig:study}
\end{figure}

Figure \ref{fig:study} summarizes the results of the blinded clinical reader study, in which two board-certified neuroradiologists evaluated full stroke-protocol image sets reconstructed from standard-of-care data and from retrospectively accelerated data, by a factor of 2, reconstructed using DPMs trained with the proposed approach.

The first reader evaluated image sets from all 80 subjects. For structural delineation, the mean scores [\text{standard}, \text{accelerated}] for gray matter, white matter, and ventricles were [4.8, 4.9], [4.8, 4.9], and [4.8, 4.9], respectively. No statistically significant difference was observed for gray matter delineation, while the proposed accelerated reconstruction achieved statistically significant improvements for white matter and ventricular delineation. For image quality metrics, mean scores [\text{standard}, \text{accelerated}] for SNR, contrast, sharpness, artifacts, and overall image quality were [4.2, 4.7], [4.7, 4.8], [4.7, 4.8], [4.1, 4.3], and [4.5, 4.8], respectively. The accelerated reconstruction demonstrated statistically significant improvements in SNR, sharpness, artifact level, and overall image quality, while no statistically significant difference was observed for contrast.

The second reader evaluated image sets from 21 subjects. For structural delineation, mean scores [\text{standard}, \text{accelerated}] for gray matter, white matter, and ventricles were [4.6, 4.1], [4.6, 4.1], and [5.0, 4.9], respectively, with no statistically significant differences observed. For image quality metrics, mean scores [\text{standard}, \text{accelerated}] for SNR, contrast, sharpness, artifacts, and overall image quality were [4.9, 4.4], [4.5, 4.1], [4.4, 4.0], [4.0, 3.9], and [4.4, 4.0], respectively. In this cohort, the standard-of-care reconstruction achieved a statistically significantly higher SNR score, while no statistically significant differences were observed for the remaining image quality metrics.

Inter-rater reliability, as measured by Cohen's kappa coefficient, was fair agreement ($0.21-0.40$) for contrast, sharpness, artifacts, and overall image quality, slight agreement ($0.00-0.20$) for SNR, white matter, and grey matter, and poor agreement ($<0.00$) for ventricles

\subsection{Experiments on Prospectively Under-sampled Data}
\begin{figure}[t]
    \centering
    \centerline{\includegraphics[width=\linewidth]{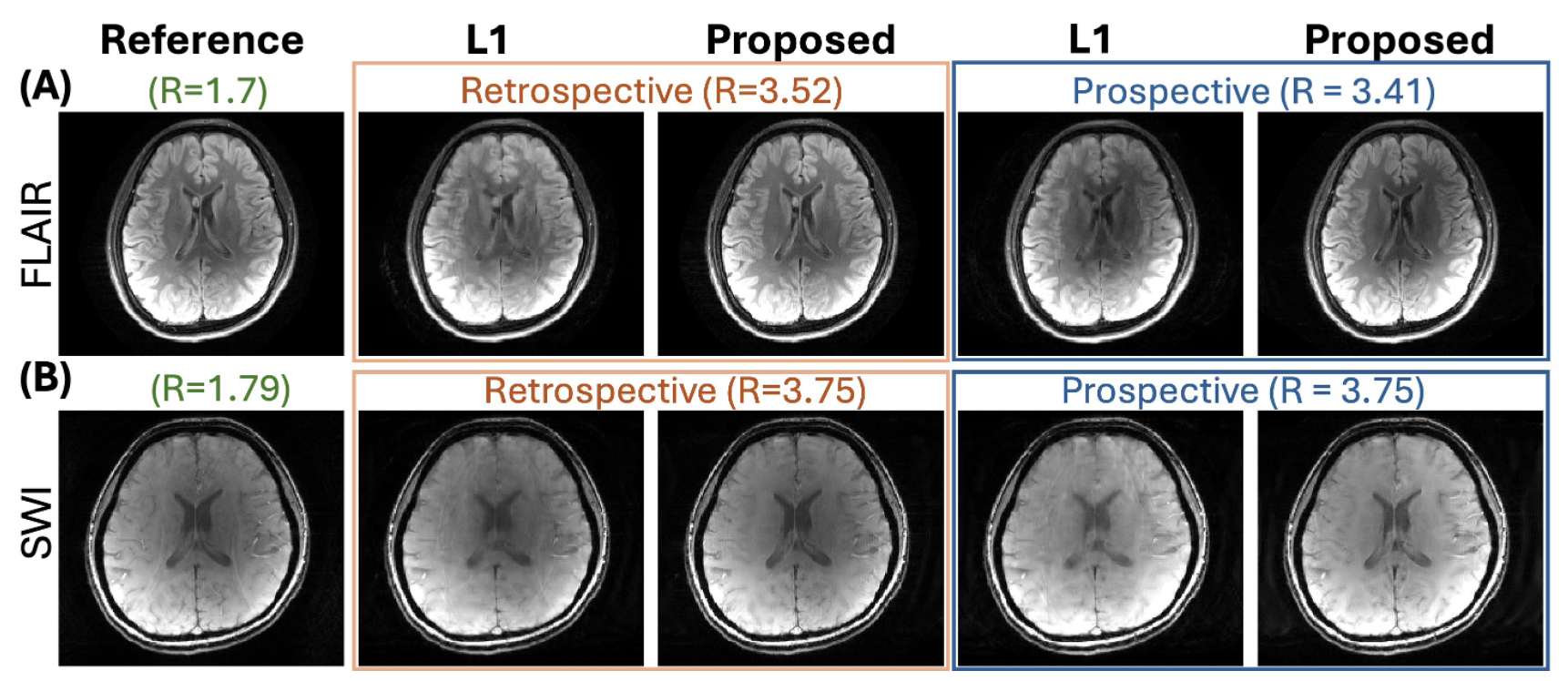}}
    \caption{The proposed DPM, pre-trained on fastMRI data and fine-tuned on FLAIR or SWI data from 20 subjects, is applied to reconstruct both retrospectively and prospectively under-sampled clinical stroke data. For each contrast, the fully sampled reference, retrospective reconstructions, and prospective reconstructions are displayed along with their effective acceleration factors. Across both contrasts, the proposed method qualitatively reduces reconstruction artifacts relative to the $\ell_1$ baseline, and achieves comparable image quality between retrospectively and prospectively under-sampled acquisitions}
    \label{fig:prospective}
\end{figure}

Figure \ref{fig:prospective} displays the model pre-trained with $T_1$, $T_2$, and post-contrast $T_1$ fastMRI and fine-tuned on FLAIR or SWI data from 20 subjects applied to reconstruct retrospectively and prospectively under-sampled FLAIR and SWI data. Comparison to an L1-wavelet baseline shows that the model qualitatively reduces reconstruction artifacts. In addition, the reconstructions from the retrospectively and prospectively under-sampled acquisitions show similar image quality.

\section{Discussion}
We proposed a simple and effective training strategy for diffusion probabilistic models (DPMs) that enables high-quality MRI reconstruction in data-constrained settings. Drawing inspiration from the foundation model paradigm, the proposed approach combines large-scale pre-training on diverse external MRI datasets followed by targeted fine-tuning on small, fully-sampled datasets from the application of interest, with an astute selection of hyperparameters. Controlled experiments on FastMRI demonstrate that DPMs trained with our strategy achieved reconstruction quality comparable to models trained on substantially larger target specific datasets. We further validated the approach on our clinically motivated stroke MRI application, where the proposed method generalized across multiple contrasts within the standard stroke imaging protocol. In a blinded clinical reader study, images reconstructed from data retrospectively accelerated by an additional factor of two and fine-tuned using only 20 fully sampled subjects were rated as non-inferior to standard-of-care reconstructions for both structural delineation and image quality metrics, 

In the controlled fastMRI FLAIR experiments, we observed a clear trade-off between insufficient adaptation and overfitting during fine-tuning. As shown in Figures \ref{fig:flaireval_quant} and \ref{fig:flaireval_qual}, omitting fine-tuning resulted in reconstructions that were poorly adapted to the target contrast, leading to suboptimal image quality. In contrast, excessive fine-tuning, either through an increased number of epochs or an overly aggressive learning rate, degraded performance, consistent with overfitting to the limited size of the target dataset. Similar trends were observed when fine-tuning to different contrasts in the clinical stroke protocol, where the optimal fine-tuning configuration varied across target datasets. Together, these results highlight that effective transfer of diffusion models to data-limited MRI reconstruction tasks requires careful selection of fine-tuning hyperparameters to balance contrast adaptation against overfitting or catastrophic forgetting \cite{kirkpatrick2017overcoming}.

The fastMRI setting further enabled a direct comparison of the proposed training strategy against alternative approaches with access to substantially larger amounts of target-domain FLAIR data (Figure \ref{fig:method_comparisons}). Methods 1–3 incorporated FLAIR data from all 344 subjects during training and therefore represent upper-bound baselines that rely on extensive target-specific supervision. Despite being fine-tuned using FLAIR data from only 20 subjects, the proposed approach (Method 4) achieved reconstruction quality comparable to these upper-bound baselines. In contrast, Methods 5 and 6, which either trained exclusively on the limited 20-subject FLAIR dataset or jointly on the large external dataset and the 20 subject FLAIR dataset, exhibited degraded performance relative to the proposed method. These results indicate that large-scale pre-training on external data is beneficial, and that sequential fine-tuning on the target domain is more effective than joint training when target-domain data are scarce.

Since Algorithm \ref{alg:DPS} represents an approximate posterior sampling procedure, fastMRI experiments in Figure 5 also characterized the effect of the data consistency weighting $\zeta$ on reconstruction performance. Higher acceleration rates corresponded to a larger optimal $\zeta$. We suspect that less total data in the data consistency term results in a lower overall value relative to the prior term, so a larger weight needs to be placed on the consistency term with less available k-space data.

Training DPMs for our limited stroke dataset motivated the proposed approach. Figures \ref{fig:swi_example}, \ref{fig:dwi_example}, and \ref{fig:protocol_comparison} demonstrate that the proposed training strategy effectively fine-tunes DPMs, with just data from 20 patients, to the various contrasts used in the stroke MRI protocol, including FLAIR, MPRAGE, SWI, and DWI. 

Results from the blinded clinical reader study (Figure \ref{fig:study}) demonstrate that images reconstructed using the proposed DPM-based approach from $2\times$ less data are non-inferior to the standard-of-care reconstructions for both structural delineation and image quality metrics. For the first reader, who evaluated all 80 subjects, the proposed approach achieved statistically significant, slightly higher scores than the standard-of-care in several categories, including SNR, artifact level, and overall image quality. In contrast, the second reader, who evaluated a smaller subset of 21 subjects, tended to rate the standard-of-care images slightly higher across most metrics. However, no statistically significant differences were observed for this reader except for SNR, where the standard-of-care achieved higher scores.

Importantly, despite these inter-reader differences, mean scores for both reconstruction approaches remained consistently high for both readers, with most metrics averaging near or above a score of 4, corresponding to good image quality. This suggests that, while individual reader preferences and scoring tendencies may differ, both the proposed and standard-of-care reconstructions produced clinically acceptable images across the evaluated stroke protocols. The observed improvement in SNR for the proposed approach in the first reader may be attributable to the denoising effect of posterior sampling guided by a learned score function, while the reduced artifact scores may reflect the exclusion of motion-corrupted k-space lines when reconstructing from fewer measurements. 

Most experiments in this work relied on retrospective under-sampling, while the prospective evaluation was limited to data acquired from a healthy volunteer. Future work will therefore focus on validating the proposed approach using prospectively under-sampled clinical stroke MRI acquisitions to more fully assess performance. In addition, future reader studies will evaluate the impact of the proposed reconstructions on downstream clinical interpretation and decision-making relative to standard-of-care images. Finally, the current posterior sampling implementation incurs substantially longer reconstruction times per slice compared to conventional parallel imaging and end-to-end learning–based methods. Future work will investigate more efficient implementations and improved posterior sampling strategies to reduce reconstruction time and facilitate practical clinical deployment.

\section{Acknowledgment}
This work was supported in part by NSF CCF-2239687 (CAREER), NSF IFML 2019844, and JCCO fellowship.

\bibliography{references}

\end{document}